\input{aipcheck}

\documentclass[
    ,final            % use final for the camera ready runs
  ]
  {aipproc}

\layoutstyle{6x9}

\begin{document}

\title{Neon and Chemical Fractionation Trends in Late-type Stellar Atmospheres}


\author{David Garc\'{\i}a-Alvarez}{
  address={Instituto de Astrof\'{\i}sica de Canarias, E-38205 La Laguna, Tenerife, 
  Spain},altaddress={GTC CALP, E-38712 San Antonio, La Palma, Spain, david.garcia@gtc.iac.es}
}

\author{Jeremy J. Drake}{
  address={Harvard-Smithsonian Center for Astrophysics, 60 Garden Street, Cambridge, MA 02138, USA}
}

\author{Paola Testa$^{**}$}{address={}
   % additional visiting address
}

\begin{abstract}
A survey of Ne, O and Fe coronal abundances culled from the recent
literature for about 60 late-type stars confirms that the Ne/O ratio
of stellar outer atmospheres is about two times the value 
recently recommended
by Asplund et al.  The mean Ne/O remains flat from the most active
stars down to at least intermediate activity levels ($-5<L_X/L_{bol}<-2$), 
with some evidence for a decline toward the
lowest activity levels sampled.  The abundances surveyed are all based
on emission measure distribution analyses and the mean Ne/O is about
0.1~dex lower than that found from line ratios in the seminal study of
mostly active stars by Drake \& Testa (2005), but is within the systematic
uncertainties of that study.  We also confirm a pattern of strongly
decreasing Fe/O with increasing stellar activity.  The observed
abundance patterns are reminiscent of the recent finding of a
dependence of the solar Ne/O and Fe/O ratios on active region plasma
temperature and indicate a universal fractionation process is at work.
The firm saturation in stellar Ne/O at higher activity levels combined
with variability in the solar coronal Ne/O leads us to suggest that Ne
is generally depleted in the solar outer atmosphere and photospheric
values are reflected in active stellar coronae.  The solution to the
recent solar model problem would then appear to lie in a combination
of the Asplund et al. (2005) O abundance downward revision being too large,
and the Ne abundance being underestimated for the Sun by about a
factor of 2. 
 
\end{abstract}

\maketitle

%%%%%%%%%%%%%%%%%%%%%%%%%%%%%%%%%%%%%%%%%%%%
%% MAINMATTER
%%%%%%%%%%%%%%%%%%%%%%%%%%%%%%%%%%%%%%%%%%%%

\section{Introduction}

The interior structure of the Sun can be studied with great accuracy
using observations of its oscillations, similar to seismology of the
Earth.  Precise agreement between helioseismological measurements and
predictions of theoretical solar models has been a triumph of modern
astrophysics \cite{Bahcall05}. However, a recent downward revision by
25-35\%\ of the solar abundances of light elements such as C, N, O and
Ne \cite{Asplund05} has broken this accordance: models adopting the
new abundances incorrectly predict the depth of the convection zone,
the depth profiles of sound speed and density, and the helium
abundance \cite{Bahcall05}. The discrepancies are far beyond the
uncertainties in either the data or the model
predictions \cite{Bahcall05b}.

\citet{Drake05a} reported on neon abundances relative to
oxygen measured in a sample of nearby solar-like stars from their
X-ray spectra.  The results were very similar for all stars and
substantially larger than the recommended solar value of Ne/O=0.15 by
number \cite{Anders89,Asplund05}. If the Ne/O abundance
in these stars is adopted for the Sun the models are brought back into
agreement with helioseismology \cite{Antia05}. However, the stellar
sample used by \citet{Drake05a} for their abundance
analysis contains mostly active stars and binary systems, which are
not comparable to the Sun in terms of activity and tend to exhibit
different outer atmosphere chemical fractionation patterns.  Hence it
is also necessary to determine the Ne/O ratio in relatively inactive
stars that are comparable to the Sun.  In this work we analyse a
sample three times larger than that of \citet{Drake05a},
including about 20 low activity stars ($\log L_{\rm X}/L_{\rm
bol}<-5$).

%%%%%%%%%%%%%%%%%%%%%%%%%%%%%%%%%%%%%%%%%%%%
%% Sample figure:
%%
%% The option [height=...] scales the picture to the given height,
%% without it it would be printed at its nominal size
%%%%%%%%%%%%%%%%%%%%%%%%%%%%%%%%%%%%%%%%%%%%

\begin{figure}
  \includegraphics[height=.5\textheight]{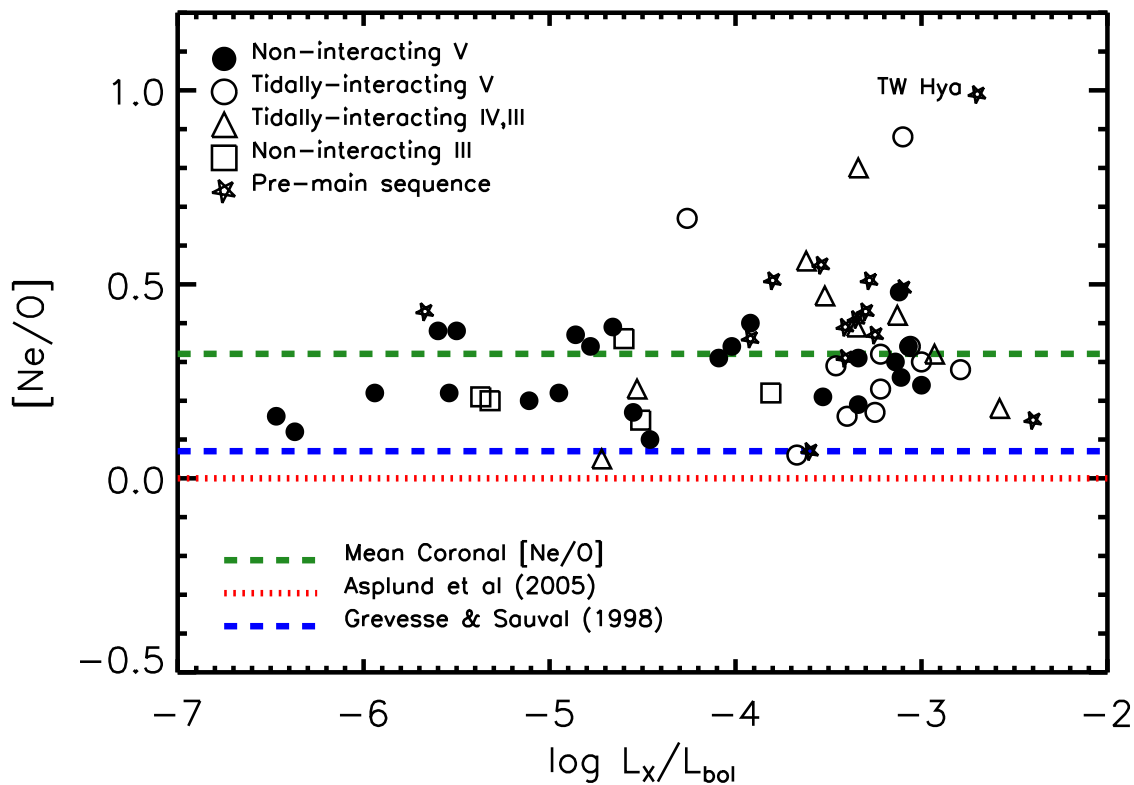}
  \caption{The coronal Ne/O abundance ratio for the stellar sample
expressed relative to the solar photospheric value of 
\citet{Asplund05}
(n(Ne)/n(O)=0.15 by number).  The highest value found is for the T
Tauri star TW Hya and was been interpretted by \citet{Drake05b} 
in terms of gain depletion of metals in gas emitting X-rays
from an accretion shock.  The mean [Ne/O] illustrated was computed
omitting this point.}
\end{figure}

\begin{figure}
  \includegraphics[height=.5\textheight]{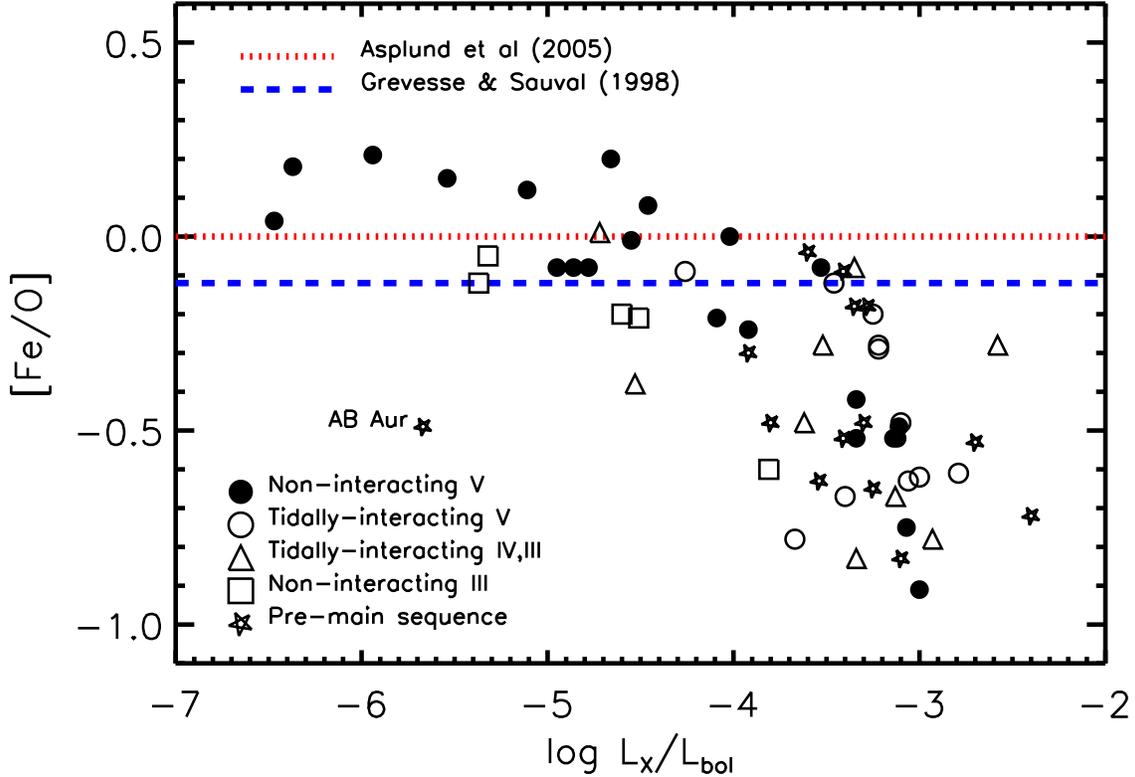}
  \caption{The coronal Fe/O abundance ratio expressed relative to
the \citet{Asplund05} solar photospheric mixture in the usual
  spectroscopic logarithmic bracket notation for the stellar
sample.  The Herbig Ae pre-main sequence star AB Aurigae is the only
deviant point from the well-defined trend of decreasing Fe/O ratio
with increasing stellar activity.}
\end{figure}

\section{Observations and Data Analysis}

Our sample is based on analyses of high resolution {\it Chandra} and
{\it XMM-Newton} spectra culled from the recent literature for about
60 late-type stars, including dwarfs, giants, tidally-interacting
binaries and pre-main sequence stars. It covers a wide range of
activity regimes ($-2 < \log L_{\rm X}/L_{\rm bol} < -7$) and spectral
types (mid-F to mid-M).  We examined the coronal Fe/O and Ne/O ratios
for these stars using the abundances derived from differential
emission measure distribution analyses performed in the various
literature studies. In the following, the coronal Fe/O and Ne/O
abundance ratios
are expressed relative to the Asplund et al.\citet{Asplund05}
solar photospheric mixture for the stellar sample.

%%%%%%%%%%%%%%%%%%%%%%%%%%%%%%%%%%%%%%%%%%%%
%% SAMPLE TABLE
%%
%% Shows the use of \tablehead and \tablenote
%% macros
%%%%%%%%%%%%%%%%%%%%%%%%%%%%%%%%%%%%%%%%%%%%
\section{Discussion and Conclusion}

%We present a survey of Ne, O and Fe coronal abundances based on
%differential emission measure distribution analyses of high resolution
%{\it Chandra} and {\it XMM-Newton} spectra culled from the recent
%literature for about 60 late-type stars, including dwarfs, giants,
%tidally-interacting binaries and pre-main sequence stars.

The mean Ne/O of our stellar sample is about twice the value recently 
recommended for
the solar photosphere \cite{Asplund05} and the average of
estimates for the solar corona, and about 0.1 dex lower than the mean
estimated by \citet{Drake05a} from line ratios for a sample of
mostly active stars, but is within the systematic uncertainties of the
latter study (see Fig.1).

There is no significant dependence of coronal Ne/O on the stellar
activity indicators $L_X/L_{bol}$ or Rossby number, though the lowest
activity stars of the sample present some evidence of a decline in
Ne/O for $L_X/L_{bol}< 6$ and solar-like activity, as has also been
suggested in a recent study of the same stars \cite{Robrade08}.
For higher activities Ne/O, for dwarfs is statistically consistant with
a constant value [Ne/O]=0.3 with observational scatter of about 0.1
dex.

In contrast to Ne/O, the Fe/O ratios vary by more than order of
magnitude with a clear trend of decreasing Fe/O with increasing
activity (see Fig.2). Low to intermediate activity stars with $L_X/L_{bol}<
-4.5$ tend to have enhanced Fe/O compared with photospheric
expectations (quasi-solar for most of the stars in the sample)
indicative of a solar-like FIP effect.  At higher activity levels,
Fe/O exhibits a strong decline with increasing activity in all the
types of stars sampled, including pre-main sequence stars.

The observed abundance patterns are reminiscent of evidence for a
dependence of the solar Ne/O and Fe/O ratios on active region plasma
temperature \cite{Drake08} and indicate a universal fractionation
process is at work.  The apparently constant stellar Ne/O over a wide
range in activity level combined with variability observed in the
solar coronal Ne/O leads us to suggest that Ne is generally depleted
in the solar outer atmosphere, whereas photospheric values are
reflected in more active stellar coronae.

If this is the case, the solution to the recent "solar model problem"
- solar models using the \citet{Asplund05} composition containing too little metal opacity to match helioseismic observations - is a
combination of the \citet{Asplund05} O abundance downward revision
being too large, and the Ne abundance being underestimated for the Sun
by about a factor of 2. 

%%%%%%%%%%%%%%%%%%%%%%%%%%%%%%%%%%%%%%%%%%%%%%%%
%% BACKMATTER
%%%%%%%%%%%%%%%%%%%%%%%%%%%%%%%%%%%%%%%%%%%%%%%%

\begin{theacknowledgments}
DGA thanks the finantial support by the Spanish Ministry of Science through
project AYA 2007-63881. JJD was funded by NASA contract
NAS8-39073 to the CXC during the course of this research.

\end{theacknowledgments}



\bibliographystyle{aipproc}   % if natbib is available

%\bibliography{sample}

\IfFileExists{\jobname.bbl}{}
 {\typeout{}
  \typeout{******************************************}
  \typeout{** Please run "bibtex \jobname" to optain}
  \typeout{** the bibliography and then re-run LaTeX}
  \typeout{** twice to fix the references!}
  \typeout{******************************************}
  \typeout{}
 }

\end{document}